\documentclass[%
reprint,
superscriptaddress,
nofootinbib,
 amsmath,amssymb,
]{revtex4-1}

\pdfoutput=1
\usepackage{float}
\usepackage{xcolor}
\usepackage{graphicx}
\usepackage{dcolumn}
\usepackage{bm}
\usepackage[colorlinks=true, linkcolor=blue, citecolor=blue, urlcolor=blue]{hyperref}
\newcommand{\ZIB}{Zuse Institute Berlin, 14195 Berlin, Germany}
\newcommand{\JCM}{JCMwave GmbH, 14050 Berlin, Germany}
\newcommand{\OHIO}{Department of Physics and Astronomy, Ohio University, Athens, Ohio 45701, United States}
\newcommand{\OPTIQUE}{LP2N, Institut d'Optique Graduate School, CNRS, Univ. Bordeaux, 33400 Talence, France}

\usepackage{graphicx}
\usepackage{booktabs}
\usepackage{tabularx}

\newcommand{\mytoprule}{\specialrule{0.1em}{0.2em}{0.2em}}
\newcommand{\mymidrule}{\specialrule{0.1em}{0.2em}{0.2em}}
\newcommand{\mybottomrule}{\specialrule{0.1em}{0.2em}{0.2em}}

\begin{document}

\title{Hot electron generation through near-field excitation \\ of plasmonic nanoresonators}

\author{Felix~Binkowski}
\affiliation{\ZIB}
\author{Tong~Wu}
\affiliation{\OPTIQUE}
\author{Philippe~Lalanne}
\affiliation{\OPTIQUE}
\author{Sven~Burger}
\affiliation{\ZIB}
\affiliation{\JCM}
\author{Alexander~O.~Govorov}
\affiliation{\OHIO}

\begin{abstract}
\vspace{0.1cm}
We theoretically study hot electron generation through the emission of a dipole source coupled to
a nanoresonator on a metal surface. In our hybrid approach, we solve the time-harmonic Maxwell's equations
numerically and apply a quantum model to predict the efficiency of hot electron generation.
Strongly confined electromagnetic fields and the strong enhancement of hot electron generation
at the metal surface are predicted and are further interpreted with the theory of quasinormal modes. 
In the investigated nanoresonator setup, both the emitting source and the acceptor resonator are localized
in the same volume, and this configuration looks promising to achieve high efficiencies of hot electron generation.
By comparing with the efficiency calculated in the absence of the plasmonic nanoresonator, that is,
the dipole source is located near a flat, unstructured metal surface, we show that the effective excitation of the modes
of the nanoresonator boosts the generation efficiency of energetic charge carriers. The proposed scheme
can be used in tip-based spectroscopies and other optoelectronic applications.
\end{abstract}
\maketitle

\section{\label{sec:intro}Introduction}
Light-matter interactions in metal nanostructures can be strongly enhanced by
plasmonic resonance effects \cite{Barnes_2003,Novotny_NatPhot_2011}.
Hot electron generation, which attracted significant attention in recent
years \cite{Linic_2011_NatMater,Brongersma_2015_NatNano,Hartland_2017_ACSEnLett,Kim_2018_NatCHem,Wu_2018_Nanoscale,DuChene_2018_NanoLett,Pensa_2019_NanoLett,Mascaretti_2020_JApplPhys},
is one important effect resulting from the absorption of plasmons by metal surfaces.
With this effect, visible light can be harvested and 
its energy can be transferred to an adjacent semiconductor, 
where the energy can then be used for photocatalytic processes \cite{Zhang_ChemRev_2018}.
The impact of morphology and materials on local field enhancement and 
hot electron generation is typically investigated in setups  
with illumination from the far field, e.g., solar illumination and other
macroscopic illumination settings \cite{Harutyunyan_2015_NatNano,Sykes_2017_NatCommun,Negrin-Montecelo_2020_ACSEnLett}. 
However, there are also various types of localized light sources accessible, such as
plasmonic tips, single molecules, quantum wells, or quantum dots \cite{Hecht_2000_JChemPhys_SNOM,Anger_2006_PRL,Senellart_2017},
which have so far not been considered for the generation of excited charge carriers.
\footnote{This work has been accepted for publication:\\
F. Binkowski et al., ACS Photonics (2021).\\
DOI: \href{https://doi.org/10.1021/acsphotonics.1c00231}{10.1021/acsphotonics.1c00231}}

The efficiency of hot electron generation in metal nanostructures depends
on the magnitude of the electric fields in the vicinity of the nanostructures \cite{Hartland_2017_ACSEnLett}.
Nanofabrication technologies allow fabrication of plasmonic nanoresonators of various
shape and characteristic size well below $100\,\textrm{nm}$ \cite{Lindquist_2012},
which enables light confinements at the nanometre scale:
The plasmonic resonances of the deep-subwavelength resonators can be efficiently excited by localized emitters
resulting in highly localized electromagnetic fields at the metal surfaces \cite{Liu_2009_PRL,Giannini_PlasmonNanoantennas_2011}.
For the design and optimization of nanophotonic devices based on emitter-resonator excitations,
modal approaches are a common theoretical tool \cite{Lalanne_QNMReview_2018}.
The localized surface plasmon resonances of the systems, which are quasinormal modes
(QNMs) \cite{Zworski_Resonances_1999,Lalanne_QNMReview_2018}, are electromagnetic field solutions
to the time-harmonic source-free Maxwell's equations. The corresponding resonance problems are
solved numerically \cite{Lalanne_QNM_Benchmark_2018}, and the solutions allow to obtain insights into the 
physical properties of the nanophotonic devices.

In this work, we investigate hot electron generation with a localized emitter
placed in the near field of a metal nanostructure.
In particular, we numerically study a circular nanogroove 
resonator on a silver surface with a characteristic size of $\sim$ $40\,\textrm{nm}$
and compare the efficiency of hot electron generation
in the presence and absence of the nanoresonator.
We compute and analyze the hot electron generation
with a quantum model assisted by full-wave simulations and further investigate
the impact of geometrical parameters. We numerically demonstrate
that the excited localized resonance of the nanoresonator leads to an
enhancement of the hot electron generation efficiency of
more than one order of magnitude compared to the flat surface.

\begin{figure*}[]
\includegraphics[width=0.7\textwidth]{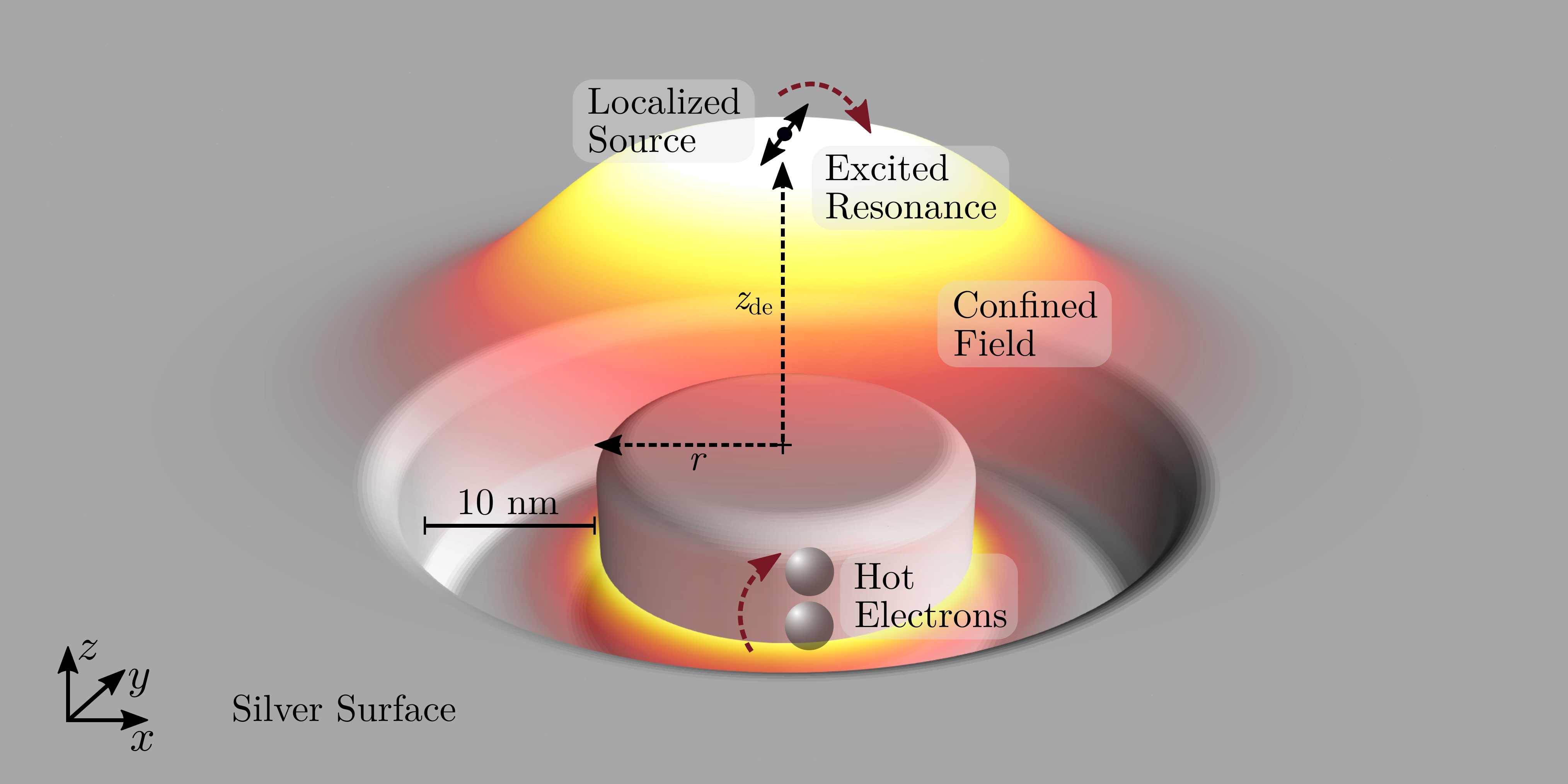}
\caption{\label{fig1} Circular nanogroove resonator with radius $r$ on a silver
surface interacting with a localized emitter placed
at the dipole-to-surface distance $z_\mathrm{de}$.
The sketched electric field intensity $|\tilde{\mathbf{E}}|^2$ corresponds to an excited
localized surface plasmon resonance.
Placing a dipole emitter close to the metal surface leads to hot electron generation.
The coupling of the emitter with the resonance yields high electric field values
localized at the nanogroove, which enhances the efficiency of hot electron generation.}
\vspace{-0.2cm}
\end{figure*}

\vspace{-0.2cm}
\section{Excitation of plasmonic resonances with localized emitters}
\label{sec2}
\subsection{Theoretical background and numerical methods}
In nano-optics, in the steady-state regime, the electric fields $\mathbf{E}(\mathbf{r},\omega_0) \in \mathbb{C}^3$
resulting from a source field are solutions to the time-harmonic Maxwell's equations in second-order form,
\begin{align}
	\nabla \hspace{-0.05cm} \times \hspace{-0.05cm} \mu^{-1} \nabla \hspace{-0.05cm} \times \hspace{-0.05cm} \mathbf{E}(\mathbf{r},\omega_0) \hspace{-0.05cm} - \hspace{-0.05cm}
	\omega_0^2\epsilon(\mathbf{r},\omega_0) \mathbf{E}(\mathbf{r},\omega_0) \hspace{-0.05cm} = \hspace{-0.05cm}
	i\omega_0\mathbf{J}(\mathbf{r}), \label{maxwell}
\end{align}
where $\omega_0\in\mathbb{R}$ is the angular frequency,
$\mathbf{r}$ is the spatial position, and $\mathbf{J}(\mathbf{r})\in \mathbb{C}^3$ is 
the electric current density corresponding to the source. 
The source field for a localized source can be modeled by a dipole source 
$\mathbf{J}(\mathbf{r}) = \mathbf{j}\delta(\mathbf{r}-\mathbf{r}')$,  
where $ \delta(\mathbf{r}-\mathbf{r}')$ is the delta distribution, $\mathbf{r}'$
is the position of the emitter, and $\mathbf{j}$ is the dipole amplitude vector.
In the optical regime, the permeability tensor $\mu$ typically equals the vacuum permeability $\mu_0$.
The permittivity tensor $\epsilon(\mathbf{r},\omega_0)$ describes the spatial
distribution of material and the material dispersion.

\begin{table}[]
	\begin{tabularx}{0.43\textwidth}{ccc} \mytoprule
		\hspace{0.25cm}$k$\hspace{0.25cm}
		&\hspace{1cm} $\Omega_k\,[\mathrm{eV}]$ \hspace{1cm}
		&\hspace{0.0cm} $\sigma_k\,[\mathrm{eV}]$ \hspace{1cm}\\
		\mymidrule
		$1$ & $3.9173-0.06084i$ &  $0.09267+0.01042i$ \\
		$2$ & $3.988-0.04605i$ &  $-0.0015342-0.062233i$ \\
		$3$ & $4.0746-0.63141i$ &  $1.4911+0.40655i$ \\
		$4$ & $4.6198-2.8279i$ &  $4.2843+4.2181i$ \\
		\mybottomrule
	\end{tabularx}
	\caption{Permittivity model for silver. Poles $\Omega_k$ and amplitudes $\sigma_k$ for the generalized Drude-Lorentz
	model \cite{Muljarov_Drude_2017} $\epsilon_\mathrm{metal,bulk}(\omega_0) = \epsilon_0(\epsilon_{\infty}-\omega_\mathrm{p}^2/(\omega_0^2+i\gamma_\mathrm{D}\omega_0))+\epsilon_0\sum_{k=1}^4 \left[ i\sigma_k/(\omega_0-\Omega_k) + i\sigma_k^*/(\omega_0+\Omega_k^*) \right]$, where $\epsilon_0$ is the vacuum permittivity,
    $\epsilon_\infty = 0.77259$, $\gamma_\mathrm{D} = 0.02228\,\mathrm{eV}$,
    and $\omega_\mathrm{p} =9.1423 \,\mathrm{eV}$.}
    \label{tab:table1}
    \vspace{-0.3cm}
\end{table}

We investigate a dipole emitter placed close to a nanoresonator.
The nanoresonator is a circular slit on a silver surface with a depth and
width of $10\,\textrm{nm}$. The structure has corner roundings with a radius of $2\, \textrm{nm}$.
Figure~\ref{fig1} shows a sketch of the geometry of the resonant system.
The dipole emitter is polarized parallel to the $z$ direction and located on axis above
the central nanocylinder at a separation distance $z_\mathrm{de}$ of the metal surface.
For clearly separating the effect of localized resonances supported by the
circular nanogroove resonator, we also investigate a second setup:
A localized source is placed at $z_\mathrm{de}$ above a flat, unstructured silver surface. 
In both cases, the permittivity of the silver material is described by 
a generalized Drude-Lorentz model resulting from
a rational fit \cite{Muljarov_Drude_2017,Garcia-Vergara_2017} to experimental
data \cite{Johnson_1972}, see Tab.~\ref{tab:table1}.
For the investigations, we choose a spectral region in the optical regime,
$200\,\mathrm{nm} \leq \lambda_0 \leq 700\,\mathrm{nm}$, with 
the wavelength $ \lambda_0=2\pi c/\omega_0$.

To numerically analyze the dipole emitter interacting with the nanoresonator and with the flat surface,
we use the finite element method.
Scattering and resonance problems are solved by applying the solver \textsc{JCMsuite} \cite{Pomplum_NanoopticFEM_2007}.
The solver employs a subtraction field approach for localized sources, 
adaptive meshing, higher order polynomial ansatz functions,
and allows to exploit the rotational symmetry of the geometry \cite{Schneider_2018}.

\begin{figure}[]
\includegraphics[width=0.485\textwidth]{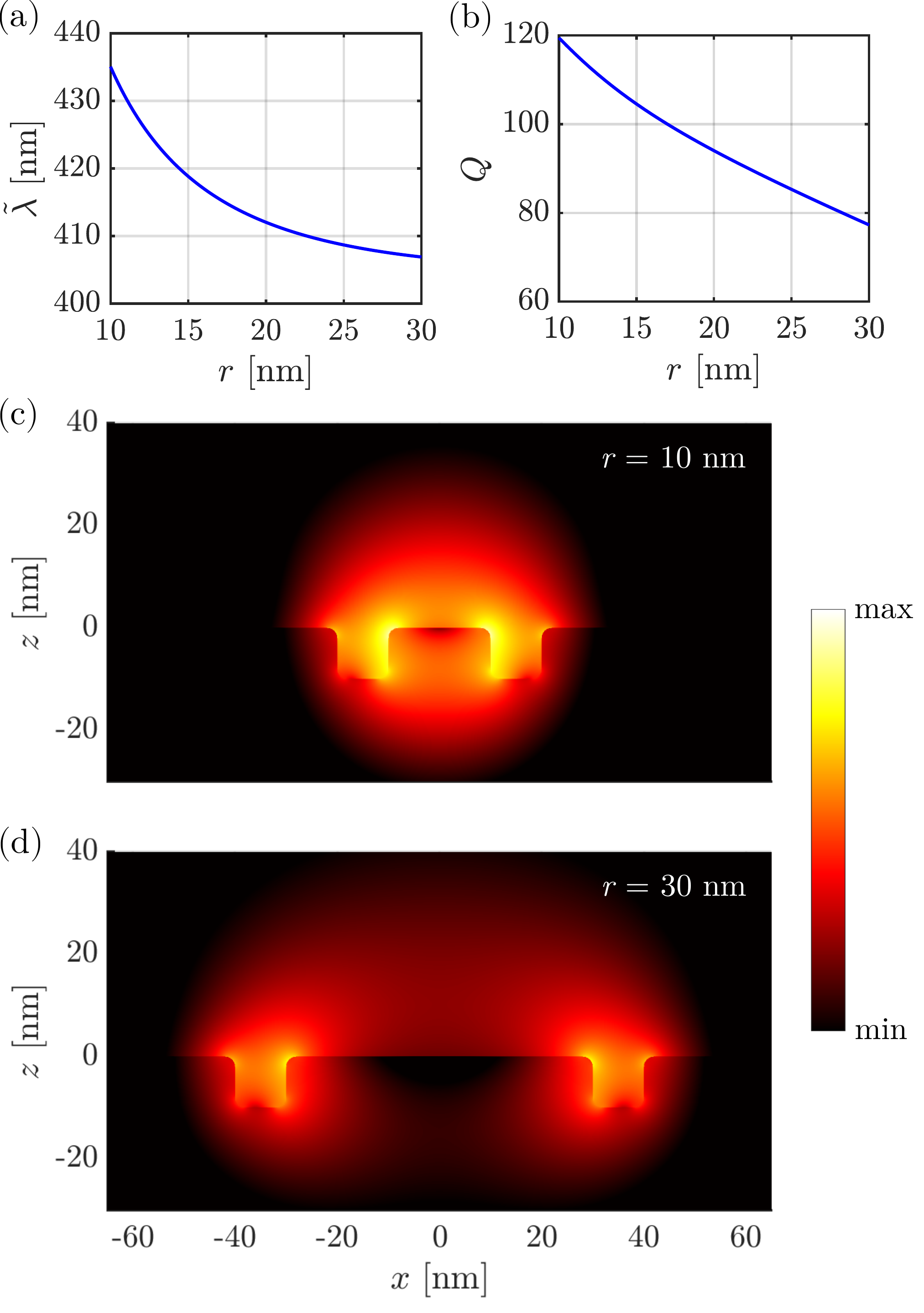}
\caption{\label{fig2}
Simulations of the circular nanogroove resonator supporting
one dominant localized resonance in the spectral region of visible light.
The associated QNM and its eigenfrequency $\tilde{\omega}$ depend
on the radius $r$ of the nanoresonator.
The permittivity model $\epsilon_\mathrm{metal,bulk}$ given in Tab.~\ref{tab:table1} is used.
(a,b)~Resonance wavelength $\tilde{\lambda} = \mathrm{Re}(2\pi c/\tilde{\omega})$ and
quality factor $Q$ of the dominant QNM, respectively. (c)~Log-plot (a.u.) of the electric field
intensity $|\tilde{\mathbf{E}}|^2$ corresponding to the dominant QNM of the nanoresonator with $r=10\,\textrm{nm}$.
The QNM is normalized \cite{Sauvan_QNMexpansionPurcell_2013} such that 
$\int_\Omega \left[ \tilde{\mathbf{E}}\cdot \frac{\partial \omega \epsilon}{\partial \omega} \tilde{\mathbf{E}} - \mu_0 \tilde{\mathbf{H}}\cdot \tilde{\mathbf{H}} \right] dV = 1$, i.e., the map allows a direct estimation and visual comparison of the interaction strength of the mode with point-like unpolarized dipoles.
The corresponding eigenfrequency is $\tilde{\omega} = (4.330-0.018i)\times10^{15}\,\textrm{s}^{-1}$ 
and the resonance wavelength is $\tilde{\lambda} = 435\,\mathrm{nm}$.
(d)~Log-plot of the electric field intensity of the normalized QNM
corresponding to the circular nanogroove resonator with $r=30\,\textrm{nm}$.}
\vspace{-0.4cm}
\end{figure}

\vspace{-0.1cm}
\subsection{Quasinormal mode analysis}
When a localized emitter is placed close to a nanostructure,
then the optical properties of the system are determined by
its underlying resonances. Localized surface plasmon
resonances, which are QNMs of the system, are one important resonance phenomena.
Figure~\ref{fig1} contains a sketch of a QNM of the nanoresonator which is investigated in this study.
QNMs are solutions to Eq.~\eqref{maxwell} with outgoing wave conditions and
without a source field, i.e., $\mathbf{J}(\mathbf{r})= 0$.
We denote the electric and magnetic field distributions of a QNM by $\tilde{\mathbf{E}}(\mathbf{r})$
and $\tilde{\mathbf{H}}(\mathbf{r})$, respectively. 
The QNMs are characterized by complex eigenfrequencies $\tilde{\omega} \in \mathbb{C}$
with negative imaginary parts. The quality factor $Q$ of a resonance,
\begin{align}
Q=\frac{\textrm{Re}(\tilde{\omega})}{-2\,\textrm{Im}(\tilde{\omega})}, \nonumber
\end{align}
describes its spectral confinement and quantifies
the relation between the stored and the dissipated electromagnetic field energy.
In the following section, we investigate how hot electron generation can be increased 
by the excitation of localized resonances.
The physical intuition behind this effect is the following:
When a localized source radiating at the frequency $\omega_0$ efficiently couples
to a localized resonance, i.e., it is spectrally ($\omega_0 \approx \mathrm{Re}(\tilde{\omega})$) and
spatially matched with the resonance,
then a large electric field $\mathbf{E}(\omega_0,\mathbf{r})$ around
the nanoresonator can be induced by the source.
At the resonance frequency $\omega_0 = \mathrm{Re}(\tilde{\omega})$, the induced field
intensity $|{\mathbf{E}}(\omega_0,\mathbf{r})|^2$ is proportional to $Q^2$,
which can significantly enhance the hot electron generation. 
Note that $|{\mathbf{E}}(\omega_0,\mathbf{r})|^2$ is also proportional
to $(\mathrm{Re}(1/\tilde{V}))^2$, where $\tilde{V}$ is the mode volume \cite{Sauvan_QNMexpansionPurcell_2013}
describing the spatial confinement of the electromagnetic field of a resonance.

In the optical regime,
the circular nanogroove resonator sketched in Fig.~\ref{fig1} supports one dominant localized resonance.
The resonance wavelength $\tilde{\lambda} = \mathrm{Re}(2\pi c/\tilde{\omega})$ 
decreases with an increasing circular slit radius $r$, see Fig.~\ref{fig2}(a).
Figure~\ref{fig2}(b) shows $Q$, depending on $r$, where $Q = 120$ can be observed for $r=10\,\textrm{nm}$.
Note that, for smaller radii, due to the decreasing radiation loss,
the quality factor would increase further. However, we restrict the investigations to $r \geq 10\,\mathrm{nm}$.
Figure~\ref{fig2}(c) shows the electric field intensity of the dominant resonance for $r=10\,\textrm{nm}$.
The resonance is strongly localized at the circular slit and is characterized by high electric field values inside 
and close to the metal.
Figure~\ref{fig2}(d) shows the electric field intensity of the dominant resonance for $r=30\,\textrm{nm}$.
It can be observed that, in comparison to the resonance for $r=10\,\textrm{nm}$, the electric field intensity
becomes smaller at the metal surface. The ratio between stored and dissipated electromagnetic field energy
decreases with an increasing radius. For the following investigations, we consider the
circular nanogroove resonator shown in Fig.~\ref{fig2}(c), which has a radius of $r=10\,\textrm{nm}$ and a quality factor of $Q = 120$.

\vspace{-0.1cm}
\subsection{Dipole emission and absorption}
To quantify the interaction of the circular nanogroove resonator with a dipole emitter close to the resonator,
we investigate the total power emitted by the dipole, which is also called dipole emission.
The dipole emission can be computed by
\begin{align}
    p_\mathrm{de}(\omega_0)= - \frac{1}{2} \mathrm{Re}\left(\mathbf{E}^*(\mathbf{r}',\omega_0) \cdot \mathbf{j}\right), \nonumber
\end{align}
where $\mathbf{E}^*(\mathbf{r},\omega_0)$ is the complex conjugate of the electric field,
$\mathbf{r}'$ is the position of the emitter, and $\mathbf{j}$ is the dipole amplitude vector.
The electric field $\mathbf{E}(\mathbf{r},\omega_0)$ is computed by solving Eq.~\eqref{maxwell} with a dipole source.

\begin{figure}[]
\includegraphics[width=0.485\textwidth]{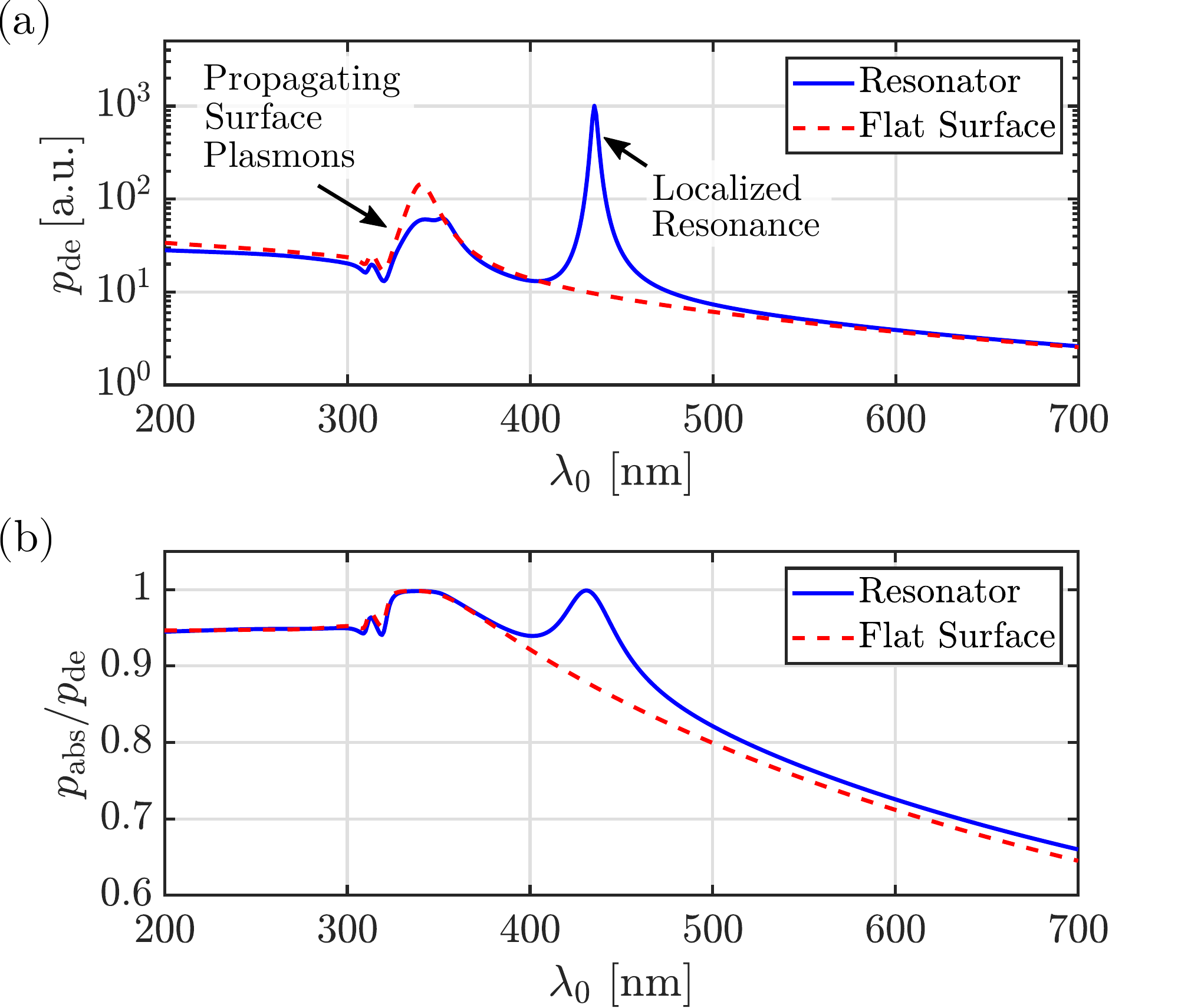}
\caption{\label{fig3}
Simulations of dipole emission and normalized absorption for a localized
source placed at 
the dipole-to-surface distance $z_\mathrm{de} = 20\,\textrm{nm}$.
Investigation for
the circular nanogroove resonator with $r=10\,\mathrm{nm}$ and comparison to a flat surface.
The permittivity model $\epsilon_\mathrm{metal,bulk}$ given in Tab.~\ref{tab:table1} is used.
(a)~Dipole emission $p_\mathrm{de}$. (b)~Normalized absorption in metal $p_\textrm{abs}/p_\textrm{de}$.}
\vspace{-0.4cm}
\end{figure}

Based on the modal results from the previous subsection, we place the dipole emitter at $z_\mathrm{de} = 20\,\textrm{nm}$, 
which is in a spatial region of high electric field intensity of the dominant resonance shown in Fig.~\ref{fig2}(c). 
In this way, the localized resonance of the circular nanogroove resonator has a significant influence on the emission
properties of the dipole emitter.
Figure~\ref{fig3}(a) shows the dipole emission $p_\mathrm{de}(\lambda_0)$.
In the case of the nanoresonator, the spectrum is characterized by two significant maxima,
which are based on different resonance effects:
The dipole emitter couples to the dominant localized resonance with the resonance wavelength
$\tilde{\lambda} = 435\,\mathrm{nm}$ and it couples also to a continuum of surface plasmons,
which are propagating on the metal surface.
As expected, the propagating surface plasmons occur not only in the presence
of the nanoresonator, but also in the case of the flat surface.
Their high density of states give rise to a peak in the spectrum between
$\lambda_0 = 300\,\mathrm{nm}$ and $\lambda_0 = 400\,\mathrm{nm}$,
as indicated in Fig.~\ref{fig3}(a), where the coupling of the dipole emitter to the propagating surface plasmons
is stronger in absence of the nanoresonator.

It can be expected that, for the investigated systems,
all energy that is not radiated into the upper hemisphere is absorbed by the metal. 
Therefore, the total absorbed energy can be computed using the expression
\begin{align}
    p_\mathrm{abs}(\omega_0) = p_\mathrm{de}(\omega_0) - p_\mathrm{rad}(\omega_0). \nonumber
\end{align}
The dipole emission radiated into the upper hemisphere, $p_\mathrm{rad}(\omega_0)$, 
is computed by a near-field to far-field transformation and an
integration of the Poynting vector over the upper hemisphere.
Figure~\ref{fig3}(b) shows the absorption $p_\textrm{abs}(\lambda_0)$ normalized by the dipole
emission $p_\textrm{de}(\lambda_0)$ for $z_\mathrm{de} = 20\,\mathrm{nm}$.
It can be observed that, close to the wavelength of the localized resonance,
most of the energy is absorbed.
As the presence of the nanoresonator increases the electromagnetic
field energy in the metal, the system with the nanoresonator
leads to a higher absorption efficiency than the system with the flat surface.

To summarize, the simulations in this subsection show that 
a localized source can efficiently excite localized resonances supported by a nanoresonator, 
as well as propagating surface plasmons on flat metal surfaces. 
In the following section, it is shown that especially excited localized resonances can 
have a significant impact on the rate at which hot electrons can be generated 
in our model system. 

\section{Hot electron generation}
\subsection{Theoretical background}
Considering quantum surface effects in plasmonics, one should start from an elegant
theory by Feibelman developed to describe a surface plasmon dispersion in metals~\cite{Feibelman_1974_PRB,Tsuei_1989_PRL}.
The so-called Feibelman's $d$-parameters characterize the dispersion and damping of the surface plasmon mode beyond the
classical electromagnetic theory. Furthermore, it was discovered that the plasmon excitations in small nanoparticles
experience an additional damping mechanism, the so-called surface-scattering decay \cite{Hartland_2011_ChemRev}.
In this quantum mechanism, collective plasmon excitations turn into hot electrons due to scattering at the
surfaces \cite{Genzel_1975_ZfPB,Kreibig_1995,Kraus_1983,Lerme_2011_JPhysChemC,Uskov_2014,Santiago_2020_ACSPhot}.
A full kinetic picture of the plasmon excitation in a nanostructure involves both low-energy ``Drude'' electrons
forming the coherent plasmon oscillation and the energetic (hot) electrons generated through
the surface-assisted Kreibig's mechanism \cite{Govorov_2013_JPhysChemC}.
The low-energy excitations, regarded above as Drude electrons, can also be derived
directly from the quasi-classical theory based on the Boltzmann equation~\cite{Ziman_1972,Tanner_2019}.
Another related work,
which should be mentioned here, is the theory of hot electron photocurrents generated at 
metal-semiconductor interfaces~\cite{Tamm_1931,Brodskii_1968,Protsenko_2012,Zhukovsky_2014}.
In our approach, we combine some of the quantum formalisms
mentioned above~\cite{Genzel_1975_ZfPB,Govorov_2013_JPhysChemC,Zhukovsky_2014,Santiago_2020_ACSPhot}
with the classical formalism of computing the electromagnetic fields at the surfaces
by solving Maxwell's equations.
The theoretical treatment below, which incorporates the surface-assisted generation of hot electrons,
is very convenient since it allows to investigate nanostructures with arbitrarily complex shapes,
in which hot-spot and shape effects determine the formation of plasmonic modes.
We note that our formalism does not include a bulk mechanism of hot electron generation due to
the electron-phonon scattering~\cite{Brown_2016}. However, such a phonon-assisted channel should not
play a dominant role in relatively small nanostructures where plasmonic mode sizes are less
than $40\,\mathrm{nm}$~\cite{Brown_2016}. In our case, the groove size of the nanostructure
is just $10\,\mathrm{nm}$, and we expect that the leading mechanism is the surface-assisted hot electron generation.
Another argument for the importance of the surface-generated hot electrons
is that those carriers are created at the surface and, therefore, can be transferred
to surface acceptor states for photochemistry or for other detection methods.

\subsection{Quantum efficiency of hot electron generation}
The rate of energy dissipation based on the generation of hot electrons at a surface
is given by \cite{Besteiro_2017}
\begin{align}
     p_\textrm{he}(\omega_0) = \frac{1}{2\pi^2}\frac{e^2E^2_\textrm{F}}{\hbar}
    \frac{1}{(\hbar\omega_0)^2} \int_S |\mathbf{E}_\mathrm{n}(\mathbf{r},\omega_0)|^2 dS, \label{HE_rate}
\end{align}
where $e$ is the elementary charge, $E_\textrm{F}$ is the Fermi energy, and $\hbar$ is the reduced Planck constant.
The normal component of the electric field
$\mathbf{E}_\mathrm{n}(\mathbf{r},\omega_0)$ is integrated over the surface $S$.
For a detailed derivation of Eq.~\eqref{HE_rate}, the reader is referred to ref \citenum{Besteiro_2017}.

The quantum dissipation $p_\textrm{he}(\omega_0)$ is based on
optically induced quantum transitions of electrons near to the surface:
The energy of photons can be transferred to the electrons because
of breaking of linear momentum conservation. This surface scattering effect can be accounted for by
a phenomenological approach for metal nanostructures~\cite{Kreibig_1995,Uskov_2014,Santiago_2020_ACSPhot}.
An additional damping mechanism with the quantum decay parameter $\gamma_\mathrm{s}$ is incorporated
in the material model,
\begin{align}
\begin{split}
    \epsilon(\omega_0) =\,\,&\epsilon_\mathrm{metal,bulk}(\omega_0) +  \epsilon_0 \frac{\omega_\mathrm{p}^2}{\omega_0(\omega_0+i\gamma_\mathrm{D})} \\
    &-\epsilon_0
    \frac{\omega_\mathrm{p}^2}{\omega_0(\omega_0 + i (\gamma_\mathrm{D} + \gamma_\mathrm{s}))},
\end{split}\label{permit_modified}
\end{align}
where $\epsilon_\mathrm{metal,bulk}(\omega_0)$ is the permittivity model for the metal bulk material, and
$\omega_\mathrm{p}$ and $\gamma_\mathrm{D}$ are the plasma frequency and the
damping constant from the Drude model, respectively, see Tab.~\ref{tab:table1}.
The quantum decay parameter $\gamma_\mathrm{s}$ describes the broadening due to the scattering of electrons at the surface.
For the calculation of $\gamma_\mathrm{s}$, we consider the total absorption power in a metal nanostructure, given by
$p_\mathrm{abs} = \mathrm{Im}\left( \epsilon(\omega_0) \right) \frac{\omega_0}{2} \int_V \mathbf{E}\cdot\mathbf{E}^*dV$,
where $\epsilon(\omega_0)$ is the permittivity model from Eq.~\eqref{permit_modified}.
It is assumed that $\omega_0^2 \gg (\gamma_\mathrm{D} + \gamma_\mathrm{s})^2$, which holds
for typical cases in nanophotonics. Applying the resulting simplification
$\mathrm{Im}\left(\epsilon(\omega_0)\right) \approx \mathrm{Im} \left(\epsilon_\mathrm{metal,bulk}(\omega_0)\right) + \epsilon_0 \frac{\omega_\mathrm{p}^2 \gamma_\mathrm{s}}{\omega_0^3}$
and splitting the absorption power $p_\mathrm{abs}$ into contributions corresponding
to bulk and surface effects yield, in particular, the surface-scattering term 
$p_\mathrm{s} = \epsilon_0 \frac{\omega_\mathrm{p}^2 \gamma_\mathrm{s}}{\omega_0^3} \frac{\omega_0}{2} \int_V \mathbf{E}\cdot\mathbf{E}^*dV$~\cite{Uskov_2014,Santiago_2020_ACSPhot}.
This term can be also computed using Eq.~\eqref{HE_rate}.
The equation $p_\mathrm{he} = p_\mathrm{s}$
can be transformed and allows to compute the quantum decay parameter $\gamma_\mathrm{s}$.
A corresponding numerical iterative approach is given by~\cite{Santiago_2020_ACSPhot}
\begin{align}
\begin{split}
    \gamma_{\mathrm{s},n} = \frac{3}{4}v_\mathrm{F} &\frac{\int_S |\mathbf{E}_\mathrm{n}(\mathbf{r},\omega_0,\gamma_{\mathrm{s},n-1})|^2 dS}{\int_V \mathbf{E}(\mathbf{r},\omega_0,\gamma_{\mathrm{s},n-1})\cdot\mathbf{E}^*(\mathbf{r},\omega_0,\gamma_{\mathrm{s},n-1}) dV}, \\ &n = 0,1,\dots,
    \end{split} \label{gamma_s}
\end{align}
where $\gamma_{\mathrm{s},0} = 0$, $v_\mathrm{F}$ is the Fermi velocity, and
the electric fields are computed by solving Eq.~\eqref{maxwell} numerically, and subsequently, they are 
integrated over the surface $S$ and the volume $V$ of the considered nanostructure.
For the computation of the electric fields within the iteration,
the material model given by  Eq.~\eqref{permit_modified} is used.
Note that, for $\gamma_{\mathrm{s},0} = 0$, we obtain $\epsilon(\omega_0) = \epsilon_\mathrm{metal,bulk}(\omega_0)$
as used for the calculations for the optical problem in the previous section.
We further note that a formalism for $\gamma_{\mathrm{s},n}$
can also be derived without the
assumption $\omega_0^2 \gg (\gamma_\mathrm{D} + \gamma_\mathrm{s})^2$~\cite{Santiago_2020_ACSPhot}.

The consideration of the quantum decay parameter $\gamma_{\mathrm{s},n}$
is equivalent of solving a self-consistent quantum-classical
formalism which fully accounts for the change of the surface response
caused by the generation of hot electrons.  
With this approach, the total power emitted by a dipole can be expressed as
\begin{align}
    p_\mathrm{de}(\omega_0) =  p_\mathrm{abs,bulk}(\omega_0) +  p_\mathrm{he}(\omega_0) +  p_\mathrm{rad}(\omega_0),
   \nonumber
\end{align}
where $p_\mathrm{abs,bulk}(\omega_0)$ is the absorption in the metal bulk.
We define the quantum efficiency of hot electron generation as the ratio
$\eta_\mathrm{he}(\omega_0) = p_\mathrm{he}(\omega_0)/p_\mathrm{de}(\omega_0)$.
This parameter describes the fraction of the dipole energy converted into hot electrons.
The efficiency of the absorption in the metal bulk and the 
radiation efficiency are defined as
$\eta_\mathrm{abs,bulk}(\omega_0) = p_\mathrm{abs,bulk}(\omega_0)/p_\mathrm{de}(\omega_0)$
and $\eta_\mathrm{rad}(\omega_0) = p_\mathrm{rad}(\omega_0)/p_\mathrm{de}(\omega_0)$, respectively.

To investigate the effect of hot electron generation for the circular nanogroove resonator,
we choose, as in the previous section, the dipole-to-surface distance $z_\mathrm{de} = 20\,\textrm{nm}$,
and solve Eq.~\eqref{maxwell} with the introduced permittivity model in Eq.~\eqref{permit_modified}.
The Fermi energy and the Fermi velocity of silver are given by $E_\textrm{F} = 5.48\,\mathrm{eV}$
and $v_\mathrm{F} = 1.39\times10^6\,\mathrm{m/s}$~\cite{Kittel_2005}, respectively.
The quantum decay parameter
$\gamma_{\mathrm{s},n}$ is obtained by the iteration in Eq.~\eqref{gamma_s}, where
the abort condition for the iteration is $|\gamma_{\mathrm{s},n}-\gamma_{\mathrm{s},n-1}|/ \gamma_{\mathrm{s},n} < 10^{-2}$.
For all simulations, with an initial value of $\gamma_{\mathrm{s},0}=0$,
this convergence condition can be achieved within a maximum of four iterations.
The electric fields $\mathbf{E}(\mathbf{r},\omega_0)$ resulting from this procedure are used
to compute $p_\mathrm{de}(\omega_0), p_\mathrm{he}(\omega_0)$, and $p_\mathrm{rad}(\omega_0)$.
To obtain the absorption in the metal bulk, we use the expression
$p_\mathrm{abs,bulk}(\omega_0) =  p_\mathrm{de}(\omega_0) -  p_\mathrm{he}(\omega_0) -  p_\mathrm{rad}(\omega_0)$.
Note that the quantum decay parameter $\gamma_{\mathrm{s},n}$ and, therefore, the quantum dissipation
$p_\mathrm{de}(\omega_0)$, depend on the size of the surface $S$ and on the size of
the volume $V$ in Eq.~\eqref{gamma_s}.
For example, for a system radiating at the wavelength of the localized resonance shown
in Fig.~\ref{fig2}(c), $p_\mathrm{he}({\lambda}_0 = 435\,\mathrm{nm})$
changes less than $1\,\%$ when the radius of the integration domains is doubled from
$1\,\mu\mathrm{m}$ to $2\,\mu\mathrm{m}$.
We choose a fixed integration radius of $2\,\mu\mathrm{m}$ for all simulations.

\begin{figure}[]
\includegraphics[width=0.485\textwidth]{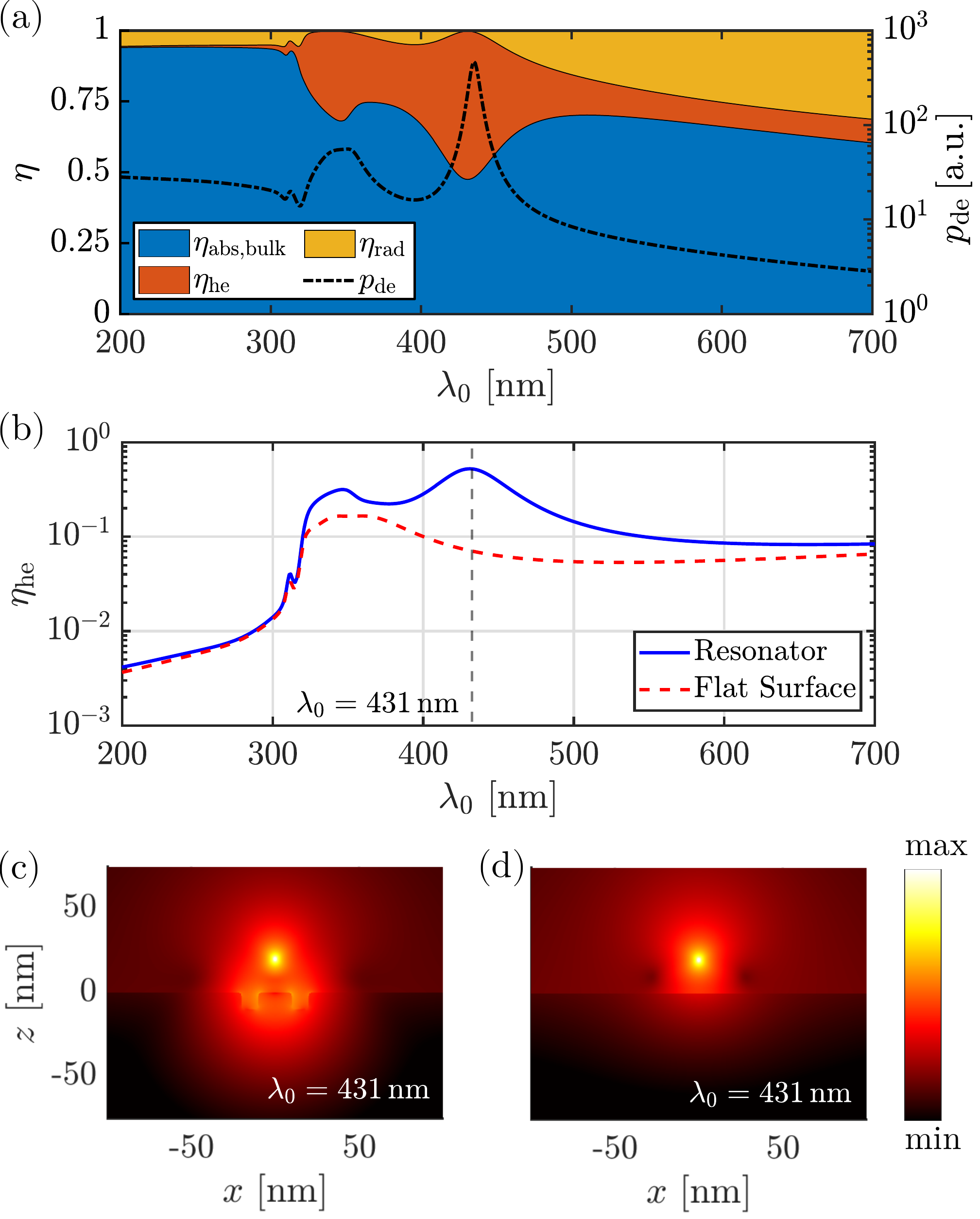}
\caption{\label{fig4} 
Simulations of hot electron generation for a localized emitter placed at
the dipole-to-surface distance $z_\mathrm{de} = 20\,\textrm{nm}$,
for the circular nanogroove resonator with $r=10\,\mathrm{nm}$ and a flat surface.
The modified permittivity function given by Eq.~\eqref{permit_modified} is used.
(a)~Left $y$ axis: Area plot for the absorption efficiency $\eta_\mathrm{abs} = p_\mathrm{abs,bulk}/p_\mathrm{de}$,
hot electron efficiency $\eta_\mathrm{he} = p_\mathrm{he}/p_\mathrm{de}$,
and radiation efficiency $\eta_\mathrm{rad} = p_\mathrm{rad}/p_\mathrm{de}$ for the nanoresonator.
Right $y$ axis: Dipole emission $p_\mathrm{de}$ for the nanoresonator.
(b)~Quantum efficiency of hot electron generation $\eta_\mathrm{he}$ for the nanoresonator and a flat surface.
(c,d)~Log-plot (a.u.) of the electric field intensity $|{\mathbf{E}}|^2$
resulting from a dipole emitter radiating at the wavelength ${\lambda}_0 = 431\,\mathrm{nm}$
for the nanoresonator and a flat surface, respectively.}
\end{figure}

Figure~\ref{fig4}(a) shows
the computed efficiencies $\eta_\mathrm{abs,bulk}(\lambda_0)$,
$\eta_\mathrm{he}(\lambda_0)$, and $\eta_\mathrm{rad}(\lambda_0)$
and the corresponding absolute values for the dipole emission $p_\mathrm{de}(\lambda_0)$.
In the full spectral range, due to the small dipole-to-surface distance, 
a large part of the power emitted by the dipole is absorbed in the metal bulk,
and only a smaller part is radiated to the upper hemisphere.
The quantum efficiency of hot electron generation $\eta_\mathrm{he}(\lambda_0)$
is significant in the spectral regions corresponding to
the localized resonance shown in Fig.~\ref{fig2}(c)
and corresponding to the propagating surface plasmons.
A comparison of the results for $p_\mathrm{de}(\lambda_0)$ in Fig.~\ref{fig4}(a) and in
Fig.~\ref{fig3}(a) shows a slight reduction of $p_\mathrm{de}(\lambda_0)$ when the quantum decay parameter 
$\gamma_{\mathrm{s},n}$ is incorporated in the material model. However, the peaks of $p_\mathrm{de}(\lambda_0)$
are still present, which demonstrates that
the optical resonance effects are the main drivers for
hot electron generation in our model system.
In both cases, with and without including the surface-scattering effect in the material model,
the maximum of the dipole emission $p_\mathrm{de}(\lambda_0)$
is located at the resonance wavelength of the localized resonance, at $\lambda_0 = 435\,\mathrm{nm}$.

Next, we compare the quantum efficiency in the presence of the nanoresonator
with the quantum efficiency for a flat, unstructured surface.
Figure~\ref{fig4}(b) shows the corresponding spectra $\eta_\mathrm{he}(\lambda_0)$.
In the case of the nanoresonator, the maximum of the quantum efficiency is located
close to the resonance wavelength of the localized resonance, and is given
by $\eta_\mathrm{he}({\lambda}_0 = 431\,\mathrm{nm}) = 0.52$,
which is about one order of magnitude larger than in case of the flat surface.
The propagating surface plasmons are responsible for another maximum
$\eta_\mathrm{he}(\lambda_0 = 346\,\mathrm{nm}) = 0.32$.
In the case of the flat surface, the quantum efficiency shows one maximum
at the wavelength
$\lambda_0 = 360\,\mathrm{nm}$, given by $\eta_\mathrm{he}(\lambda_0 = 360\,\mathrm{nm}) = 0.17$.
The spectra $\eta_\mathrm{he}(\lambda_0)$ demonstrate
that the presence of the nanoresonator 
has a significant influence on the generation of energetic charge carriers.
Figure~\ref{fig4}(c) and (d) emphasize this by showing, for the circular
nanogroove resonator and the flat surface, respectively,
the electric field intensities in the vicinity of the dipole emitter radiating at
the wavelength ${\lambda}_0 = 431\,\mathrm{nm}$,
where the quantum efficiency is maximal.
The localized source strongly excites the localized resonance of the nanoresonator,
which leads to high electric field values at the metal surface enabling enhanced hot electron generation.
Note that, close to the wavelength of the localized resonance,
the absolute values for the dipole emission $p_\mathrm{de}(\lambda_0)$
are more than one order of magnitude larger for the system with the nanoresonator
than for the system without the nanoresonator, see also Fig.~\ref{fig3}(a).

\subsection{Dependence of hot electron generation on emitter placement}
\begin{figure}[]
\includegraphics[width=0.485\textwidth]{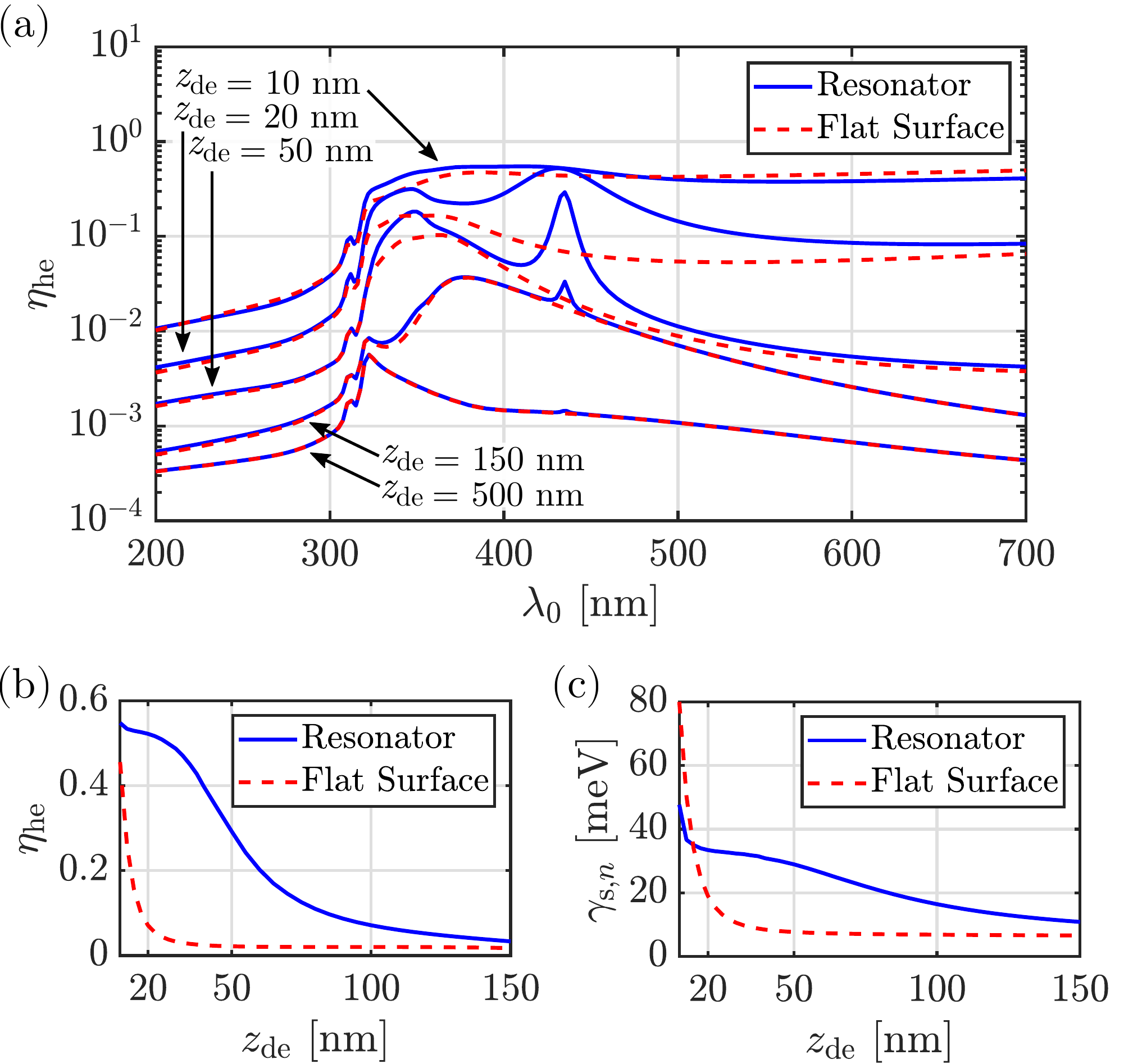}
\caption{\label{fig5} 
Simulations of hot electron generation for a localized emitter placed at different dipole-to-surface distances $z_\mathrm{de}$,
for the circular nanogroove resonator with $r=10\,\mathrm{nm}$ and a flat surface.
The modified permittivity function given by Eq.~\eqref{permit_modified} is used.
(a)~Quantum efficiency $\eta_\mathrm{he}$ as a function of emitter wavelength for various distances $z_\mathrm{de}$.
(b,c)~Quantum efficiency $\eta_\mathrm{he}$ and quantum decay parameter $\gamma_{\mathrm{s},n}$, respectively,
depending on $z_\mathrm{de}$.
The number $n$ is the last step of the  iteration in Eq.~\eqref{gamma_s}.
Note that the emitter wavelength changes as $z_\mathrm{de}$ is varied to match the spectral position of the
peak in the spectrum due to the localized resonance. The same wavelength is used for the flat surface.}
\end{figure}

Localized light sources can excite resonances that cannot be excited by illumination from the far field,
such as dark surface plasmon modes~\cite{Liu_2009_PRL} or modes
where the overlap integral with the field caused by the far-field illumination is negligible.
This allows for additional degrees of freedom in tailoring the light-matter interaction.
It can be expected that the position of the dipole emitter in our model system is a degree of
freedom that has a significant influence on the generation of excited charge carriers.
For investigating this impact, we perform simulations of the hot electron generation with various
dipole-to-surface distances. 
The corresponding results are shown in Figure~\ref{fig5}(a).
In the full spectral range, with a decreasing dipole-to-surface distance
from $z_\mathrm{de} = 500\,\mathrm{nm}$ to $z_\mathrm{de} = 10\,$nm, 
the quantum efficiency $\eta_\mathrm{he}(\lambda_0)$ strongly increases. 
The most significant effect can be observed at the 
peak in the spectrum corresponding to the localized resonance.
This can be explained through the $z_\mathrm{de}$-dependent overlap
between localized resonance and source near fields:
The resonance excitation and the resulting electromagnetic near fields increase
when the dipole-to-surface distance becomes smaller.
Note that, below $20\,\mathrm{nm}$, the efficiency at the peak does not
further increase significantly with a decrease of the distance.
This can be understood by considering that, below $20\,\mathrm{nm}$,
almost all emitted energy has already been funneled into the localized resonance,
and a further decrease of the distance does not change the electric field distribution near the metal surface.
Such a saturation of the hot electron generation efficiency can only be predicted with
self-consistent formulas, as given by Eqs.~\eqref{maxwell}, \eqref{permit_modified}, and~\eqref{gamma_s}.

Next, we investigate the behavior of the resonance-induced hot electron 
generation peak by performing a fine sampling of the dipole-to-surface distance $z_\mathrm{de}$. 
Figure~\ref{fig5}(b) shows the corresponding dependence of the quantum efficiency $\eta_\mathrm{he}$.
In the case of the nanoresonator, the quantum efficiency varies over one order of magnitude,
from $3\,\%$ to $52\,\%$,
when the distance decreases from 150\,nm to 20\,nm.
In the case of the flat surface, the quantum efficiency only increases from $2\,\%$ to $7\,\%$
when the distance decreases from  $150\,\mathrm{nm}$ to $20\,\mathrm{nm}$.

By changing the dipole-to-surface distance further, from  $z_\mathrm{de} = 20\,\mathrm{nm}$ to $z_\mathrm{de} = 10\,\mathrm{nm}$,
an additional significant effect can be observed in the case of the flat surface: The quantum efficiency
increases by more than one order of magnitude, up to $\eta_\mathrm{he} = 0.46$.
For such small distances,
high-$k$ surface plasmon polaritons can be excited~\cite{Ford_1984}. These high-$k$ surface plasmons have
a very small skin depth, which leads to strongly confined electric fields close to the metal surface.
This strong effect is not observed when the nanoresonator is present because,
in this case, the response is fully dominated by the localized resonance and the energy does
not funnel into high-$k$ surface plasmons.
As a result, when $z_\mathrm{de} = 10\,\mathrm{nm}$, the same order of magnitude of quantum efficiency is obtained in the
presence and in the absence of the nanoresonator.
Figure~\ref{fig5}(c) shows the dependence of the quantum decay parameter $\gamma_{\mathrm{s},n}$
on the distance $z_\mathrm{de}$. The quantum dissipation at the surface and the absorption power
in the metal bulk are related to the nominator and the denominator in 
Eq.~\eqref{gamma_s}, respectively. For decreasing dipole-to-surface distances,
the quantum dissipation increases faster than the absorption in the metal bulk leading to an increase~of~$\gamma_{\mathrm{s},n}$.

Along with the additional broadening of the plasmon resonance described by $\gamma_{\mathrm{s},n}$,
the surface-assisted hot electron generation processes create a peculiar, nonthermal energy distribution
of excited electrons inside a driven plasmonic nanocrystal~\cite{Besteiro_2017,Santiago_2020_ACSPhot}.
The computed shapes of nonthermal energy distributions in a nanocrystal can be found in the
refs \citenum{Santiago_2020_ACSPhot} and \citenum{Besteiro_2017}. The intraband hot electrons,
which we study here, are generated near the surface, and their spectral generation rate has
a nearly-flat distribution in the energy interval $E_\mathrm{F}<E<E_\mathrm{F}+\hbar \omega_0$.
Because of the frequent electron-electron collisions, the high-energy hot electrons experience
fast energy relaxation. Therefore, the resulting numbers of hot electrons in the steady states of plasmonic nanostructures
are always limited. However, those  hot electrons, when generated, have a good chance to be injected
into electronic acceptor states at the 
surface~\cite{Linic_2011_NatMater, Kim_2018_NatCHem, Wu_2018_Nanoscale, Mascaretti_2020_JApplPhys, Sausa-Castillo_2016, Baturina_2019}.
These electronic acceptors can be in the form of semiconductor
clusters ($\mathrm{TiO}_2$)~\cite{Sausa-Castillo_2016, Baturina_2019} or adsorbed molecular
species~\cite{Wu_2018_Nanoscale, Mascaretti_2020_JApplPhys}.  Consequently, the injected long-lived hot electrons
can cause chemical reactions in a solution~\cite{Kim_2018_NatCHem, Wu_2018_Nanoscale, Mascaretti_2020_JApplPhys}
or surface growth~\cite{Khorashad_2020}. Such chemical and shape transformations can be observed in experiments.

Based on the above results, we expect that in potential experimental setups that use hot electron generation by 
localized sources and nanostructured samples, the significant spectral dependence
and position dependence of the generation rate can provide strong experimental signatures 
and thus can provide guidelines for settings with high-efficiency hot electron generation.

\section{Conclusions}
We analyzed the hot electron generation due to the emission of light by a localized emitter placed in the near
field of a metal nanoresonator with electromagnetic field calculations and an approximate quantum model. 
For a resonant nanostructure on the metal surface, enhanced hot electron generation was observed.
This enhancement is based on a plasmonic resonance excited by the emitter. We showed that,
for a specific nanoresonator on a silver surface, the quantum efficiency is about one order
of magnitude larger than the quantum efficiency of hot electron generation in the case of a flat silver surface.
We further demonstrated a strong spectral and position dependence of the hot electron generation on
the placement of the emitter. In particular, the resonance significantly favors these effects.

The physical reason behind the efficient energy
conversion in our system is that both the exciting source and the nanoresonator have the same dimensionality:
They are zero-dimensional and, therefore, highly localized. 
Experimentally, a zero-dimensional source of radiation
is the key element in the field of tip-enhanced spectroscopies, which includes
scanning near field optical microscopy (SNOM) \cite{Heinzelmann_1994,Chen_2012_Nature},
hot electron nanoscopy \cite{Giugni_2017}, and hot electron tunneling settings \cite{Wang_2015}.
In tip-driven spectroscopy, electromagnetic fields and the related hot electron excitation
processes become strongly confined in small volumes, leading to a strong enhancement of light-matter interaction. 
Our approach can also be used to investigate coatings with quantum dots or other emitters 
on resonance-supporting surfaces. 
The presented study provides a theoretical background for hot electron generation with localized light sources.
\vspace{-0.1cm}

\section*{Acknowledgments}
F.B.~and S.B.~acknowledge funding
by the Deutsche Forschungsgemeinschaft (DFG, German Research Foundation) under Germany's Excellence Strategy - The
Berlin Mathematics Research Center MATH+ (EXC-2046/1, project ID: 390685689) and by the Helmholtz Association within
the Helmholtz Excellence Network SOLARMATH (ExNet-0042-Phase-2-3), a strategic collaboration of
the DFG Excellence Cluster MATH+ and Helmholtz-Zentrum Berlin.
P.L.~acknowledges the support from the NOMOS (ANR-18CE24-0026-03) and ISQUAD (ANR-18CE47-0006-04) projects.
A.O.G.~acknowledges support by the United States-Israel Binational Science Foundation (BSF).

\end{document}